\documentclass[aps,prd,10pt,twocolumn,superscriptaddress,showpacs]{revtex4-1}

\pdfoutput=1
\usepackage{hyperref}
\usepackage{graphicx}
\usepackage{amsfonts,amsmath,amssymb,bm,bbm}
\usepackage{color}

\usepackage{soul}

\usepackage[normalem]{ulem}

\hypersetup{
    pdfnewwindow=true,      % links in new window
    colorlinks=true,       % false: boxed links; true: colored links
    linkcolor=blue,          % color of internal links
    citecolor=blue,        % color of links to bibliography
    filecolor=blue,      % color of file links
    urlcolor=blue        % color of external links
}

\begin{document}
\title{First Observation of Time Variation in the Solar-Disk Gamma-Ray Flux with Fermi}

\author{Kenny C. Y. Ng}
\email{ng.199@osu.edu}
\affiliation{Center for Cosmology and AstroParticle Physics (CCAPP), Ohio State University, Columbus, OH 43210}
\affiliation{Department of Physics, Ohio State University, Columbus, OH 43210}

\author{John F. Beacom}
\email{beacom.7@osu.edu}
\affiliation{Center for Cosmology and AstroParticle Physics (CCAPP), Ohio State University, Columbus, OH 43210}
\affiliation{Department of Physics, Ohio State University, Columbus, OH 43210}
\affiliation{Department of Astronomy, Ohio State University, Columbus, OH 43210} 

\author{Annika H. G. Peter}
\email{apeter@physics.osu.edu}
\affiliation{Center for Cosmology and AstroParticle Physics (CCAPP), Ohio State University, Columbus, OH 43210}
\affiliation{Department of Physics, Ohio State University, Columbus, OH 43210}
\affiliation{Department of Astronomy, Ohio State University, Columbus, OH 43210} 

\author{Carsten Rott}
\email{rott@skku.edu}
\affiliation{Department of Physics, Sungkyunkwan University, Suwon 440-746, Korea}

\date{7 November 2015}

\begin{abstract}

The solar disk is a bright gamma-ray source.  Surprisingly, its flux is about one order of magnitude higher than predicted.  As a first step toward understanding the physical origin of this discrepancy, we perform a new analysis in 1--100\,GeV using 6 years of public Fermi-LAT data. 
Compared to the previous analysis by the Fermi Collaboration, who analyzed 1.5 years of data and detected the solar disk in 0.1--10\,GeV, we find two new and significant results:  1.~In the 1--10\,GeV flux~(detected at $>5\sigma$), we discover a significant time variation that anticorrelates with solar activity.  2.~We detect gamma rays in 10--30\,GeV at $>5\sigma$, and in 30--100\,GeV at $> 2\sigma$.  The time variation strongly indicates that solar-disk gamma rays are induced by cosmic rays and that solar atmospheric magnetic fields play an important role.  Our results provide essential clues for understanding the underlying gamma-ray production processes, which may allow new probes of solar atmospheric magnetic fields, cosmic rays in the solar system, and possible new physics.  Finally, we show that the Sun is a promising new target for ground-based TeV gamma-ray telescopes such as HAWC and LHAASO.

\end{abstract}
 

% 95.85.Pw Gamma rays astronomical observations
% 96.60.Hv Magnetic field solar
% 96.60.Tf Electromagnetic radiation solar
% 96.60.Mz Photosphere solar
% 96.60.Q- Solar activity,
% 13.85.Qk Photons production in hadron-induced high-energy interactions
% 13.85.Tp Cosmic rays high-energy interactions
% 96.50.Vg Solar particles and photons (cosmic rays)
% 96.50.S- Cosmic rays

\maketitle

%%%%%%%%%%%%%%%%%%%%%%%%%%%%%%%%%%%%%%%%%%%%%%%%%%%%%%%%
%%%%%%%%%%%                                SECTION                                                           %%%%%%%%%
%%%%%%%%%%%%%%%%%%%%%%%%%%%%%%%%%%%%%%%%%%%%%%%%%%%%%%%%
\section{Introduction}
\label{sec:Introduction}

The Sun is well studied and understood with a broad set of messengers at different energies.  For example, the optical photon and MeV neutrino spectra confirm a detailed picture of the Sun as a middle-aged G-type main-sequence star powered by nuclear fusion~\cite{2010ARA&A..48..289G, 2013ARA&A..51...21H}.  However, the gamma-ray emission from the Sun is poorly understood.  Precision studies of the Sun at GeV energies are only now possible after the 2008 launch of the Fermi Gamma-Ray Space Telescope~(Fermi).

Naively, one does not expect the quiet Sun~(also known as the steady or the quiescent Sun) to produce an appreciable GeV gamma-ray flux.  Even though the solar atmospheric temperature rises to millions of Kelvin in the corona, it corresponds to $\lesssim$\,keV in energy.  And, although solar flares can accelerate particles non-thermally, bright flares are rare and the highest-energy gamma ray observed from a flare is only $\simeq 4$ GeV~\cite{1996A&AS..120C.299S, Ackermann:2014rma, 2014ApJ...789...20A, 2015ApJ...805L..15P}.  

There are, however, two distinct processes involving cosmic rays that guarantee the continuous production of gamma rays from the vicinity of the Sun.  The first contribution comes from the Inverse-Compton~(IC) scattering of cosmic-ray electrons and positrons with solar photons~\cite{Orlando:2006zs, Moskalenko:2006ta, Orlando:2013pza}.  The IC component appears as an extended halo~($\sim{\cal{O}}(10^{\circ})$) around the Sun.  The second contribution comes from the hadronic interaction of cosmic rays with the solar atmosphere~(photosphere and chromosphere)~\cite{1991ApJ...382..652S}.  The extent of this component has the angular size of the Sun~($\simeq 0.5^{\circ}$); we denote it~(plus any potential non-cosmic-ray contribution) as the solar-disk component.

Theoretical estimation of both components requires taking into account the effects of solar magnetic activity.  Magnetic fields carried by the solar wind modulate the fluxes of cosmic-ray particles in the solar system~\cite{Strauss:2012zza, 2013LRSP...10....5O, 2013LRSP...10....3P}.  This effect is expected to be stronger for the solar-disk component than the IC component because of the much closer approach to the Sun for the parent cosmic rays.  In addition, magnetic fields in the solar atmosphere~\cite{2006RPPh...69..563S,  2012LRSP....9....5W, 2012LRSP....9....6M} affect the solar-disk component.  Seckel et al.~\cite{1991ApJ...382..652S}~(denoted as SSG1991 in the following) showed that solar atmospheric magnetic fields could boost gamma-ray production through the magnetic reflection of the primary cosmic rays or their showers out of the Sun.  Consequently, they estimated that the Sun could be detected by space-based gamma-ray telescopes.

The first experiment to have the sensitivity to detect {quiet Sun} gamma rays was the Energetic Gamma Ray Experiment Telescope~(EGRET)~\cite{1997JGR...10214735T}.  A reanalysis of the EGRET data later reported the first detection of  {solar-disk} gamma rays, but the flux uncertainties were large~\cite{Orlando:2008uk}.
More recently, with the improved sensitivity of the Large Area Telescope~(LAT) on board Fermi, the IC and solar-disk components were each well measured at 0.1--10\,GeV in Abdo et al.~\cite{Abdo:2011xn}~(denoted as Fermi2011 in the following).  The IC component was detected out to 20$^{\circ}$ from the Sun, and was found to be consistent with theoretical expectations~\cite{Orlando:2006zs, Moskalenko:2006ta, Orlando:2013pza}.  
{Although the observed solar-disk component satisfies the theoretical upper bound derived in SSG1991~(the naive case), it is in complete disagreement with the nominal model of SSG1991, the one and only theoretical prediction:} \emph{The observed flux is about one order of magnitude higher at all energies and the spectrum shape is flatter than predicted.}
This mismatch motivates new theoretical modeling and new observational studies of the solar-disk gamma rays.  The latter is the focus of this study.

After Fermi2011, two key questions naturally surfaced concerning solar-disk gamma rays.
First, does the solar-disk gamma-ray flux have a long-duration time variation?  
In Fermi2011, after comparing to the results from Ref.~\cite{Orlando:2008uk}, it was pointed out that a significant variation of the solar-disk emission may be present. 
If such a variation is confirmed, and if it is related to the solar activity cycle, it could test the cosmic-ray origin of the gamma rays and help reveal their production mechanism.
Second, does the Sun shine in gamma rays beyond 10\,GeV?  {The last two data points from the Fermi2011 solar-disk energy spectrum suggest the spectrum might become softer at higher energy.  Interactions of cosmic rays with solar magnetic fields are energy dependent; a spectral cutoff at high energy could reveal the end of magnetic field effects on the cosmic-ray interactions.}
It is only possible to answer these questions now because of the improved statistics and long time baseline ($>6$ years) of the Fermi-LAT data set.

We aim to address these questions in this work, which is structured as follows:  In Sec.~\ref{sec:fermi}, we present our analysis and findings.  In Sec.~III, we first provide a short overview of the hadronic solar gamma-ray production by cosmic rays.  Then we discuss future prospects for both theory and observation. Seasoned readers on cosmic-ray theory can skip the overview~(Sec.~III\,A) and move on to the rest of the section.  We conclude in Sec.~\ref{sec:conclusion}.

%%%%%%%%%%%%%%%%%%%%%%%%%%%%%%%%%%%%%%%%%%%%%%%%%%%%%%%%
%%%%%%%%%%%                                SECTION                                                           %%%%%%%%%
%%%%%%%%%%%%%%%%%%%%%%%%%%%%%%%%%%%%%%%%%%%%%%%%%%%%%%%%
\section{The Sun observed using Fermi-LAT }
\label{sec:fermi}
\subsection{Outline of the Analysis}

Launched in 2008 on board Fermi, the LAT instrument is a pair-conversion gamma-ray detector sensitive to energies from about $10^{-2}$\,GeV to $10^3$\,GeV~\cite{2009ApJ...697.1071A, 2012ApJS..203....4A}.  Its large field of view allows it to survey the whole sky.  With 1.5 years of data, Fermi2011 detected the solar-disk and IC components separately in 0.1--10\,GeV.  Since then, Fermi not only collected more data, but its quality has also improved.  Fermi data are publicly available, which allows us to perform this study.

Due to the apparent motion of the Sun on the sky, one needs to trace its position continuously with time to produce a Sun-centered image.  Because we focus on the solar-disk component, all other sources of emission are treated as backgrounds.  There are two main backgrounds that need to be accounted for; both are small compared to the signal.  The first is the diffuse background that consists of astrophysical emission~(smeared due to the motion of the Sun) and the detector background.  The second background~(technically, a foreground) is the IC component in the line of sight.  {Both backgrounds can be estimated from the data. }

We follow Fermi2011 by removing data near the Galactic plane and model the diffuse background using the fake-Sun method.  In addition, we remove all the bright solar flares.  To increase photon statistics, we relax the point-source cut and moon cut used Fermi2011.  We study and take into account the possible systematics associated with this step.

To extract the solar-disk signal, we perform a likelihood analysis with the data binned in both energy and angle.  This allows us to perform a simple and conservative analysis to characterize the main features of the signal.  
The accuracy goal of this analysis is limited by the systematic uncertainty of Fermi-LAT's effective area, which is estimated to be about $10\%$~\cite{2012ApJS..203....4A}, so we ignore uncertainties that are much less than that.  
We discuss possible ways to improve the analysis in Sec.~\ref{sec:discussion}.   

\subsection{Data Selection and Cuts}
\label{sec:data_selection}

We choose our analysis energy range to be 1--100\,GeV.  Below 1\,GeV, the point spread function~(PSF) of Fermi-LAT deteriorates rapidly, making it difficult to isolate the solar-disk component~(in addition, the Fermi Collaboration is performing a dedicated analysis at low energies~\cite{Giglietto:2015}).  Above 100\,GeV, although we find 3 photons~(up to $\sim 300$\,GeV) within $1^{\circ}$ of the center of the Sun in the final photon map, it is difficult to estimate the background contribution due to the small number of photons.

%+++++++++++++++++++++++++++++FIGURE++++++++++++++++++++++++++++++++++++++++%
\begin{figure*}[t]
\includegraphics[width=8.5cm]{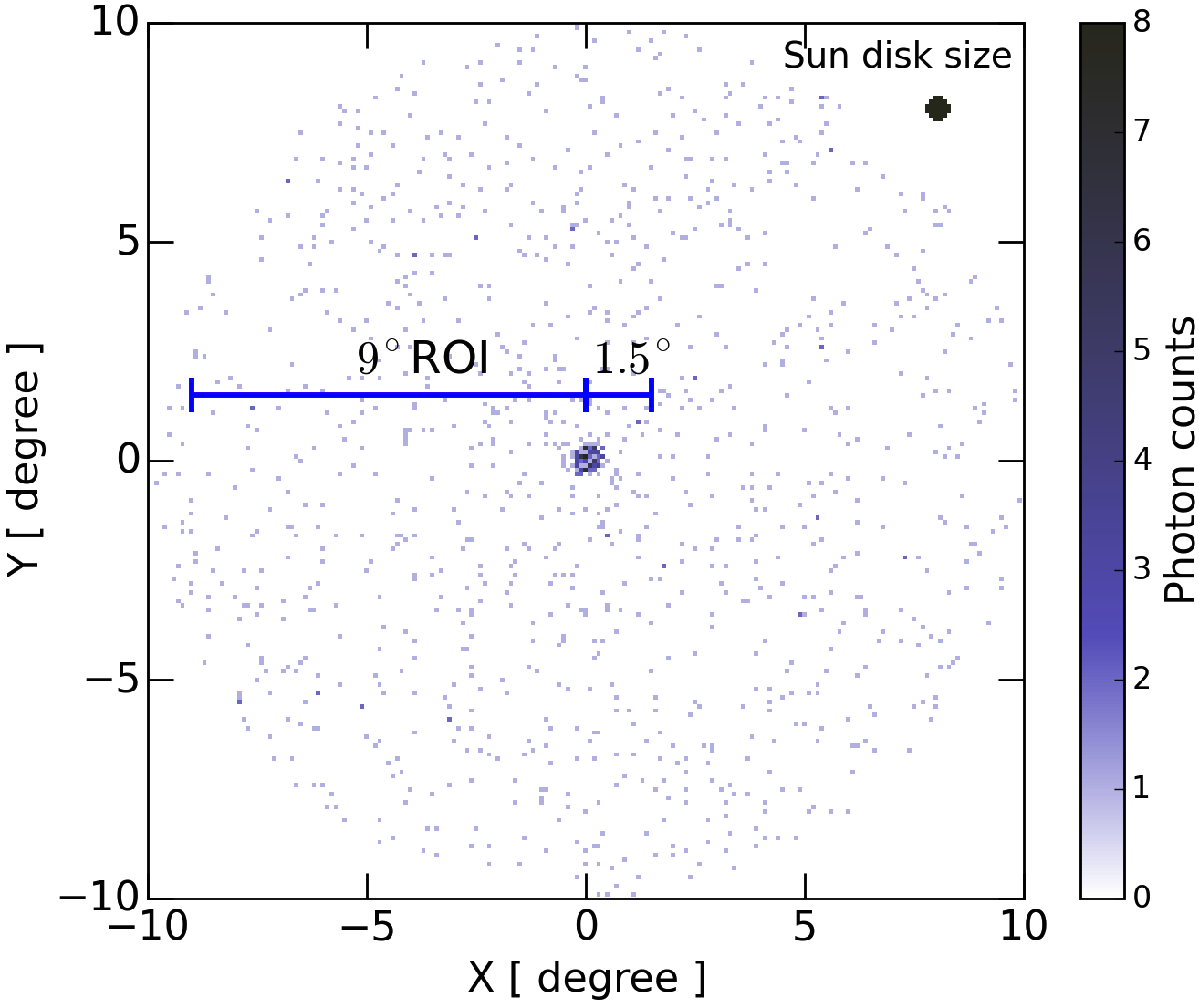}
\includegraphics[width=8.5cm]{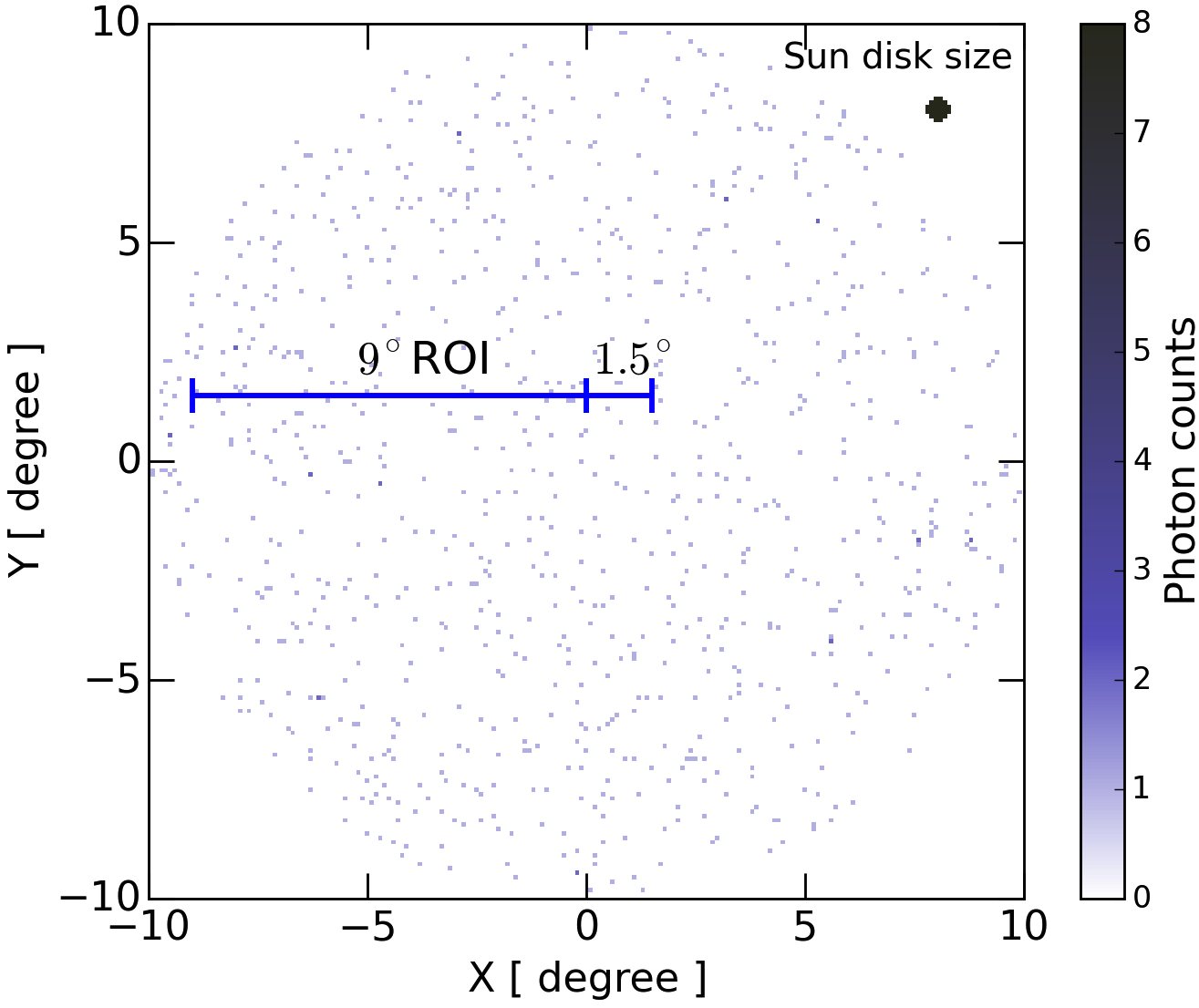}
\caption{{\bf Left:} Stacked photon counts map of the Sun ROI in 10--100\,GeV.  {\bf Right:} Same, but for a fake-Sun ROI (in this example, trailing the Sun in its path by +180 days), which is used to measure the diffuse background.  The exposures of the two ROIs differ by $\lesssim 2\%$. (Maps for $>0.1$\,GeV are shown in Fermi2011.)  Visually, the solar-disk component (comparable in extent to the size of the Sun, as marked) is obvious; that of the IC component (decreasing with angle) is more subtle.  The numbers of photons within $1.5^{\circ}$ of the center are 175 versus 19; the numbers in $1.5^{\circ}$--$9^{\circ}$ are 844 versus 710. }
\label{fig:counts_map}
\end{figure*}
%++++++++++++++++++++++++++++++++++++++++++++++++++++++++++++++++++++++++++++%

We analyze the data using the Fermi science tools version \texttt{v9r33p0}~\footnote{http://fermi.gsfc.nasa.gov/ssc/data/analysis/software/}.  We use the weekly \texttt{P7REP} data set from week \texttt{010} to week \texttt{321}, which covers from \texttt{2008-08-07} to \texttt{2014-07-31}.  (\texttt{Pass 8} data became available during the final stages of this work; we discuss this in Sec.~\ref{sec:discussion}.)
To trace the Sun's position, we divide each week into 40 identical time segments.   Because the Sun moves $\simeq 7^{\circ}$ per week, its positional drift per time segment is $\simeq 0.2^{\circ}$. {This is smaller than the diameter of the Sun~($\simeq 0.5^{\circ}$) and the LAT PSF  at 1\,GeV~($\simeq 1^{\circ}$).  Above 10\,GeV, the drift becomes comparable to the PSF~($\simeq 0.1^{\circ}$), which we mitigate by using large angular bins in the likelihood analysis.  }

For each time segment, {we adopt the standard data selection procedure recommended by the collaboration. 
We use \texttt{gtselect} to select photons from the \texttt{SOURCE} event class and to divide the events into eight energy bins of equal logarithmic width.  We set the maximum zenith angle to be 100$^{\circ}$ to avoid photons coming from the luminous Earth limb~\cite{2009PhRvD..80l2004A, Ackermann:2014ula}. }
We select all photons within 10$^{\circ}$ of the Sun; to avoid potential edge effects, we define our region of interest~(ROI) as a 9$^\circ$-radius circle.  The photon events are filtered using \texttt{gtmktime} with the keywords \texttt{DATA\_QUAL==1}, \texttt{LAT\_CONFIG==1}, and \texttt{ABS(ROCK\_ANGLE<52)}.  The first two keywords ensure that the data quality is good enough for a point-source analysis;  {the last one requires that the spacecraft be within the range of rocking angles used during nominal sky-survey observations.}
The filtered photon events are binned into photon counts maps in equatorial coordinates using \texttt{gtbin} with a pixel size $0.1^{\circ} \times 0.1^{\circ}$.  The photon maps are stacked to construct a single map for each energy bin.  

To calculate the expected number of photons from an underlying intensity~(flux per solid angle) distribution, we obtain the exposure map using \texttt{gtltcube} and \texttt{gtexpcube2} with identical settings as for the photon maps, and using the \texttt{P7REP\_SOURCE\_V15} instrumental response function.  The flux map is obtained by dividing the stacked photon map by the stacked exposure map.  The total exposure in the ROI is about $\simeq 10^{11}\,{\rm cm^{2}\,s}$, and is spatially uniform at the $\sim 1\%$ level in 1--100\,GeV. 

To check our data selection procedures, we measure the gamma-ray flux from one of Fermi's calibration sources, the Vela pulsar, which is the brightest {steady} astrophysical gamma-ray source above 0.1\,GeV.  We repeat the same data selection procedures, except for the time segments used to trace the Sun, to obtain the photon map and exposure map.  The gamma-ray flux is estimated from the total flux within $1.5^{\circ}$ of Vela, after subtracting the background estimated from the $6^{\circ}$--$9^{\circ}$ region of the same ROI.  The flux obtained is consistent with that in Ref.~\cite{etc.:2010vw}.

%+++++++++++++++++++++++++++++FIGURE++++++++++++++++++++++++++++++++++++++++%
\begin{figure*}[t]
%\centering
\includegraphics[angle=0.0, width=17cm]{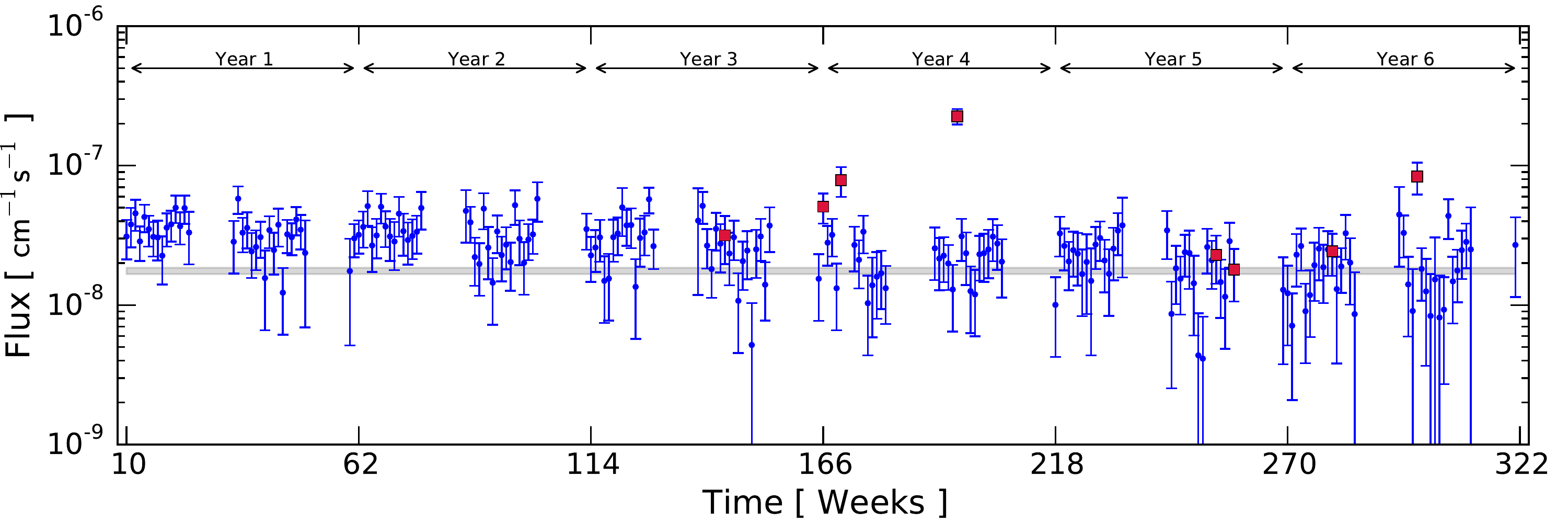}
\caption{ Total gamma-ray flux between 1\,GeV and 1.8\,GeV within 1.5$^{\circ}$ from the Sun versus time.  Each bin corresponds to one week of observation, starting from \texttt{2008-08-07} (week \texttt{010}).  Periods that coincide with a bright solar flare are labeled with red squares; these are removed from the analyses.  {The horizontal grey band shows the resulting 6-year combined flux and its uncertainty. }
} 
\label{fig:time_curve}
\end{figure*}
%+++++++++++++++++++++++++++++FIGURE++++++++++++++++++++++++++++++++++++++++%

Following Fermi2011, we remove data when $|b|<30^{\circ}$, where $b$ is the Galactic latitude.  This avoids the bright diffuse and point-source emission from the Galactic plane.  After this cut, the exposure time is reduced by $\simeq 40\%$ and the total photons by $\simeq 76\%$, consistent with the values in Fermi2011.  This cut is efficient for reducing background contamination, but is conservative because the Galactic plane emission decreases rapidly with Galactic latitude.  We discuss in detail the remaining background components in Sec.~\ref{sec:bkg_est}.

{ In Fermi2011, data are excluded whenever a known point source or the Moon is within 20$^{\circ}$ of the Sun. In order to maximize the photon counts in high energy, we relax these cuts.  Point sources are expected to increase the diffuse background by about 10\%, which has minimal effect to our solar-disk-centric analysis. The Moon should not affect our analysis because its energy spectrum falls rapidly above 1\,GeV~\cite{2012ApJ...758..140A}.  We describe in the next section in detail how we handle the inclusion of background sources in the likelihood analysis.  Imposing the point-source cut would reduce the exposure time by at least a factor of 3~(shown in Fermi2011 with 1FGL), making the high-energy analysis significantly more difficult.  (The IC component has a smaller signal-to-noise ratio.  As a result, the point-source cut is more important for an IC-centric analysis, as in Fermi2011.)}

With the goal of searching for time variations in the solar-disk flux, we pay special attention to possible time-varying sources.  The most important ones are solar flares~\cite{1996A&AS..120C.299S, Ackermann:2014rma, 2014ApJ...789...20A}.  During the period of bright solar flares, the flaring regions can emit a significant flux of gamma rays for a short period of time, thus contaminating the solar-disk signal and potentially changing the time profile of solar-disk flux.  Only a few flares are expected to matter, as solar flares are typically dim beyond a few GeV.  Another special source is the blazar 3C 279, which overlaps the coordinates of the Sun every October~\cite{2014ApJ...784..118B}.  This blazar has a flux comparable to that of the Sun and the Sun stays about a day near its location, hence it would nominally contaminate the solar-disk component at the  $\sim$1\% level.  However, when it is in a flaring state, it can temporarily be 100 times brighter~\cite{2012ApJ...754..114H, 2015arXiv150204699H}.  We check and find that the Sun was never nearby during the reported 3C 279 flares.

Figure~\ref{fig:counts_map}~(left) shows the stacked photon map in 10--100\,GeV.  It is clear from the density and the brightness of the pixels the solar disk is observed.  This is the first time that the Sun has been detected with $>10$\,GeV photons.  Compared to the map shown in Fermi2011, which was for all photons above 0.1\,GeV and is thus dominated by low-energy photons, this image is sharper due to the improved PSF at higher energies.  The right panel of Fig.~\ref{fig:counts_map} is a ``fake-Sun'' photon map, used as a background estimate, described in the next subsection. 

Figure~\ref{fig:time_curve} shows the gamma-ray flux~(1--1.8\,GeV) within 1.5$^{\circ}$ of the Sun as a function of time.  We label the time periods that contain solar flares detected by Fermi at greater than 10\,$\sigma$~\footnote{http://hesperia.gsfc.nasa.gov/fermi/lat/qlook/lat$\_$events.txt}. Some anomalously bright periods are correlated with solar flares, most notably the ones in 7 March 2012 and 25 February 2014~(week \texttt{196} and \texttt{299}).  Beyond that, we do not observe any obvious excesses. 
For consistency, all labeled periods are removed from the Sun and fake-Sun analyses.  

\subsection{Background Estimation}\label{sec:bkg_est}
\subsubsection{Diffuse Background}
Due to the motion of the Sun on the sky, all astrophysical emission is smeared to a diffuse and isotropic background.  This includes truly diffuse as well as resolved and unresolved point-source emission.  We denote this emission together with the detector background~(misidentified cosmic rays) as the diffuse background.  

We estimate the expected contribution of the diffuse background in the Sun ROI using the fake-Sun method described in Fermi2011.  We repeat identical analyses~(including all cuts) at positions where the Sun would have been $+60$, $+90$, $+180$, and $-90$ days away from the actual time.  The fake Suns traverse the same paths through the sky as the Sun, which allows us to measure the diffuse background independently. 

Figure~\ref{fig:counts_map}~(right) shows the stacked photon map in 10--100\,GeV for one of the fake Suns~(+180 days).  The Sun and fake-Sun ROIs have comparable exposures~($\lesssim 2\%$ difference).  {As a result, the small excess of photons away from the center of the Sun ROI already shows hints of the extended IC component, which becomes apparent when the angular distribution of the intensity is shown. }

The combination of four fake Suns allows us to estimate the diffuse background with better than $ 10\%$ statistical uncertainty.  However, when comparing the individual fake-Sun background estimates, we observe, at the low end of our energy range, $\simeq$ 10\% variations among the fake Suns, which is larger than their individual statistical uncertainties.  Upon closer inspection, we found that this is driven by one particularly brighter fake-Sun ROI~(+180), while the other three agree with each other at subpercent level.  We check and do not find any significantly bright periods in this fake-Sun ROI.  Therefore, this flux enhancement is likely due to one or several mild time-varying background sources, an arguably expected consequence of including point sources in the data set.  We combine the four fake-Sun ROIs to estimate the diffuse background, and mitigate the potential background variation  by adding a 10\% systematic uncertainty to the diffuse background in the likelihood analysis.  We also check our result using the background estimates without the +180 fake-Sun ROI.  The difference is miniscule.

We compare our combined fake-Sun background estimate with that from Fermi2011, and find that our background estimate is higher by $\sim$10\% at the low energy end.  Though this is consistent with systematic variation described above, it could also be explained by background sources.
The average point-source contribution to the diffuse background can be estimated using the total high-latitude~($|b|>20^{\circ}$) point-source intensity reported in the Fermi Isotropic Gamma-Ray Background analysis~(see Fig.~8 in Ref.~\cite{Ackermann:2014usa}).  Comparing this to the diffuse background in the fake-Sun ROI~(Fig.~3 in Fermi2011), point sources contribute about $10\%$ of the total diffuse background, which matches the difference seen in our fake-Sun analysis versus that in Fermi2011.  Because this extra small contribution affects both the Sun and fake-Sun ROIs, it is self-consistently modeled in the likelihood analysis. Nonetheless, we add an additional 10\% systematic uncertainty to the diffuse background in the likelihood analysis.  These systematic uncertainties~(20\% of our fake-Sun estimate) take into account all the potential systematics introduced to the diffuse background by including the point sources.  

{Lastly, the gamma-ray intensity of the fake-Sun ROIs are found to the uniform in radial direction.  This is consistent with the finding from Fermi2011, which showed that the only source of anisotropy is the Galactic plane, which we have removed. }
This angular dependence allows us to separate the diffuse background from the signal and the IC component. 

%+++++++++++++++++++++++++++++FIGURE++++++++++++++++++++++++++++++++++++++++%

\begin{figure*}[t]
\includegraphics[width=8.5cm]{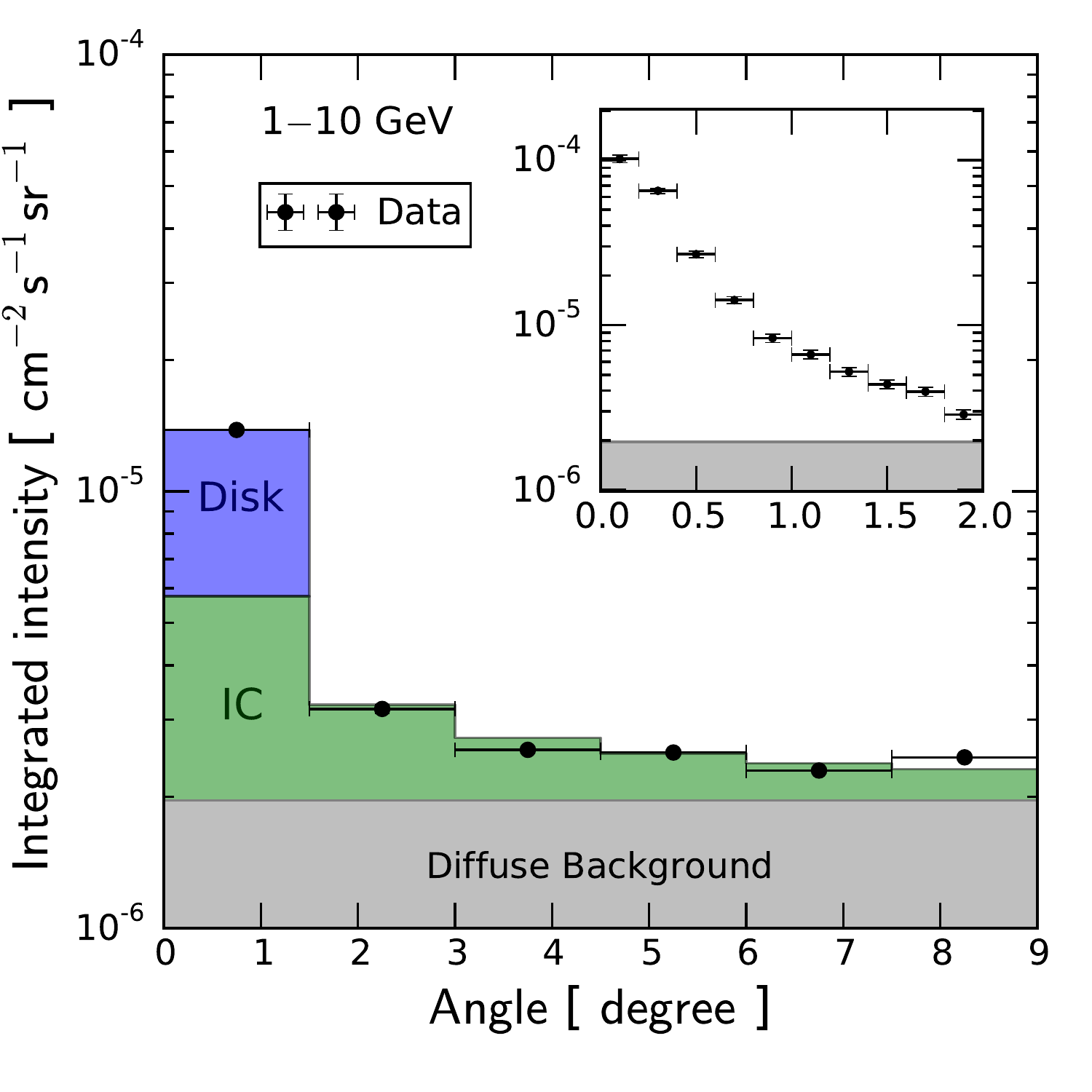}
\includegraphics[width=8.5cm]{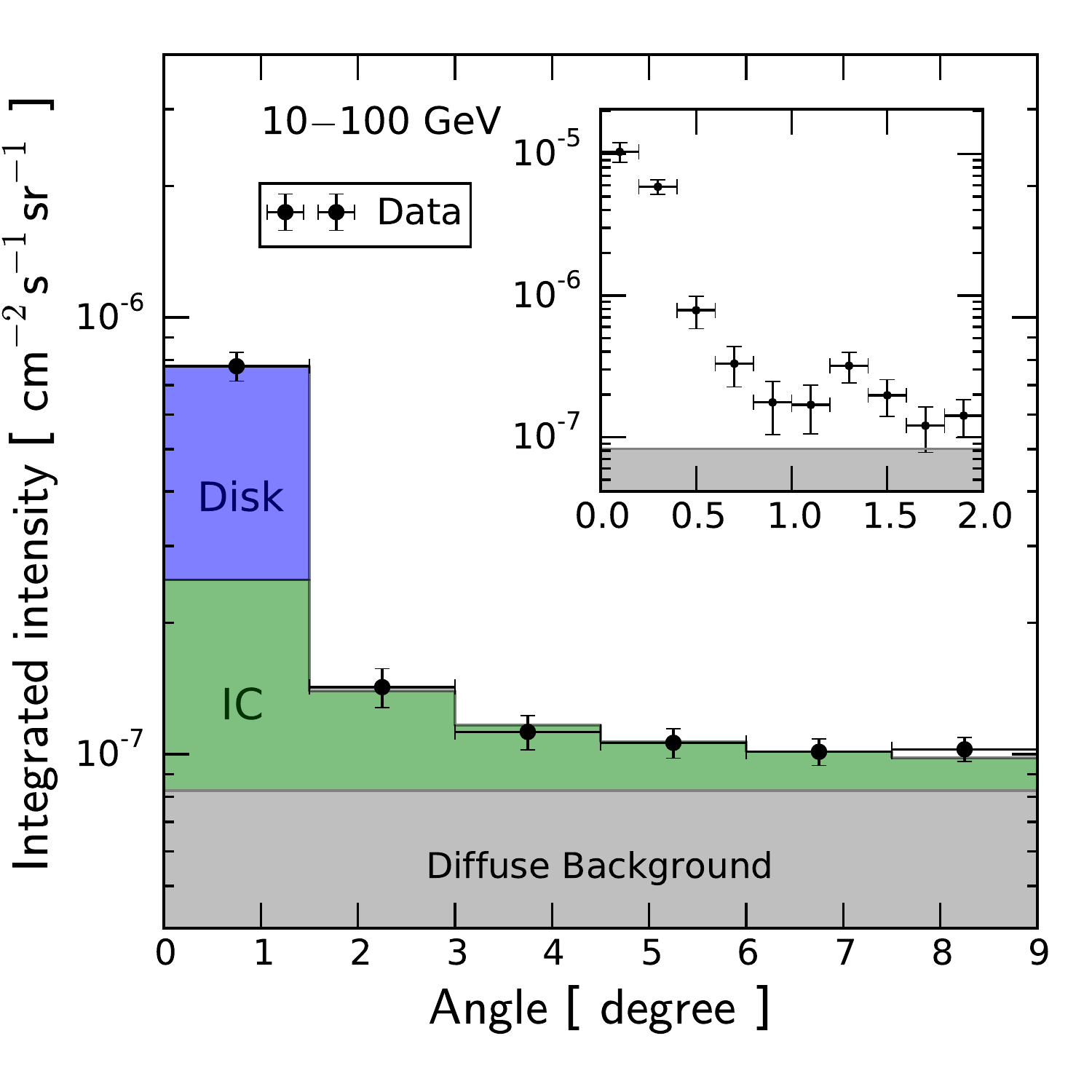}
\caption{{\bf Left:} Angular distribution of the integrated intensity from 1--10\,GeV in the Sun ROI.  Black points show the observed data with statistical uncertainties only.  Colored histograms show the fitted results for the signal and two backgrounds (the estimate of the diffuse background incorporates independent data from the fake-Sun ROIs).  The inset shows the same data with smaller angular bins, but without the two solar components (note the different vertical scale).  {\bf Right:} Same, but for 10--100\,GeV~(note the lower flux).}
\label{fig:profile}
\end{figure*}
%+++++++++++++++++++++++++++++FIGURE++++++++++++++++++++++++++++++++++++++++%

\subsubsection{Inverse Compton Emission}

In addition to the diffuse background, the extended IC component also contributes to the total emission in the Sun ROI.  We model the IC component background using its distinctive angular distribution.  Assuming the cosmic-ray electron density is homogeneous throughout the solar system, the IC component intensity is simply proportional to the column density of solar optical photons~\cite{Orlando:2006zs, Moskalenko:2006ta, Orlando:2013pza}.  This description was found to be reasonable in Fermi2011, especially for gamma-ray energies above 1\,GeV.  With this assumption, we can approximate the IC intensity as $\propto \alpha^{-1}$, where $\alpha$ is the angular distance from the Sun.  This distribution deviates from the true one~\cite{Orlando:2006zs, Moskalenko:2006ta, Orlando:2013pza} slightly at large angles, and is accurate at the $\sim 5\%$ level at the edge of our ROI.  In {the angular region of the solar disk}, the IC component is suppressed, which we take into account in the analysis~(described below).  Overall, small uncertainties of the shape of the IC component do not affect our results, as it is subdominant compared to the solar-disk emission {in the inner 1.5$^{\circ}$}.

%+++++++++++++++++++++++++++++FIGURE++++++++++++++++++++++++++++++++++++++++%
\begin{figure}[t]
%\centering
\includegraphics[angle=0.0, width=8.5cm]{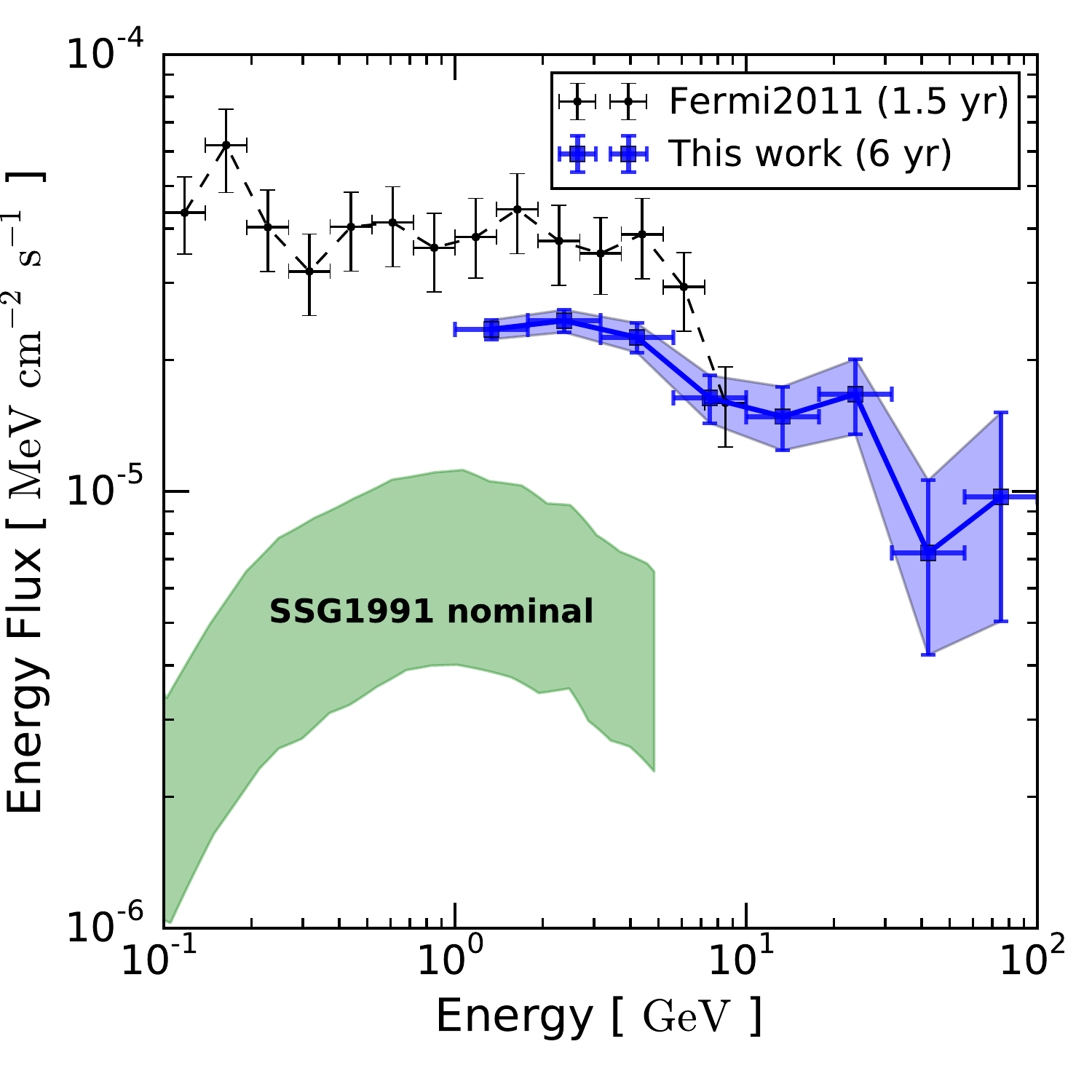}
\caption{Energy spectrum of the solar-disk flux.  Blue squares and statistical uncertainties (systematic uncertainties, not shown, are $\simeq 10\%$) are the results of our analysis with 6 years of data.  Black dots and combined statistical and systematic uncertainties are the Fermi2011 results with 1.5 years of data.  {The green band shows the predicted flux range from the SSG1991 nominal model.}  }
\label{fig:data_result}
\end{figure}
%+++++++++++++++++++++++++++++FIGURE++++++++++++++++++++++++++++++++++++++++%

%+++++++++++++++++++++++++++++FIGURE++++++++++++++++++++++++++++++++++++++++%
\begin{figure}[t]
%\centering
\includegraphics[angle=0.0, width=8.5cm]{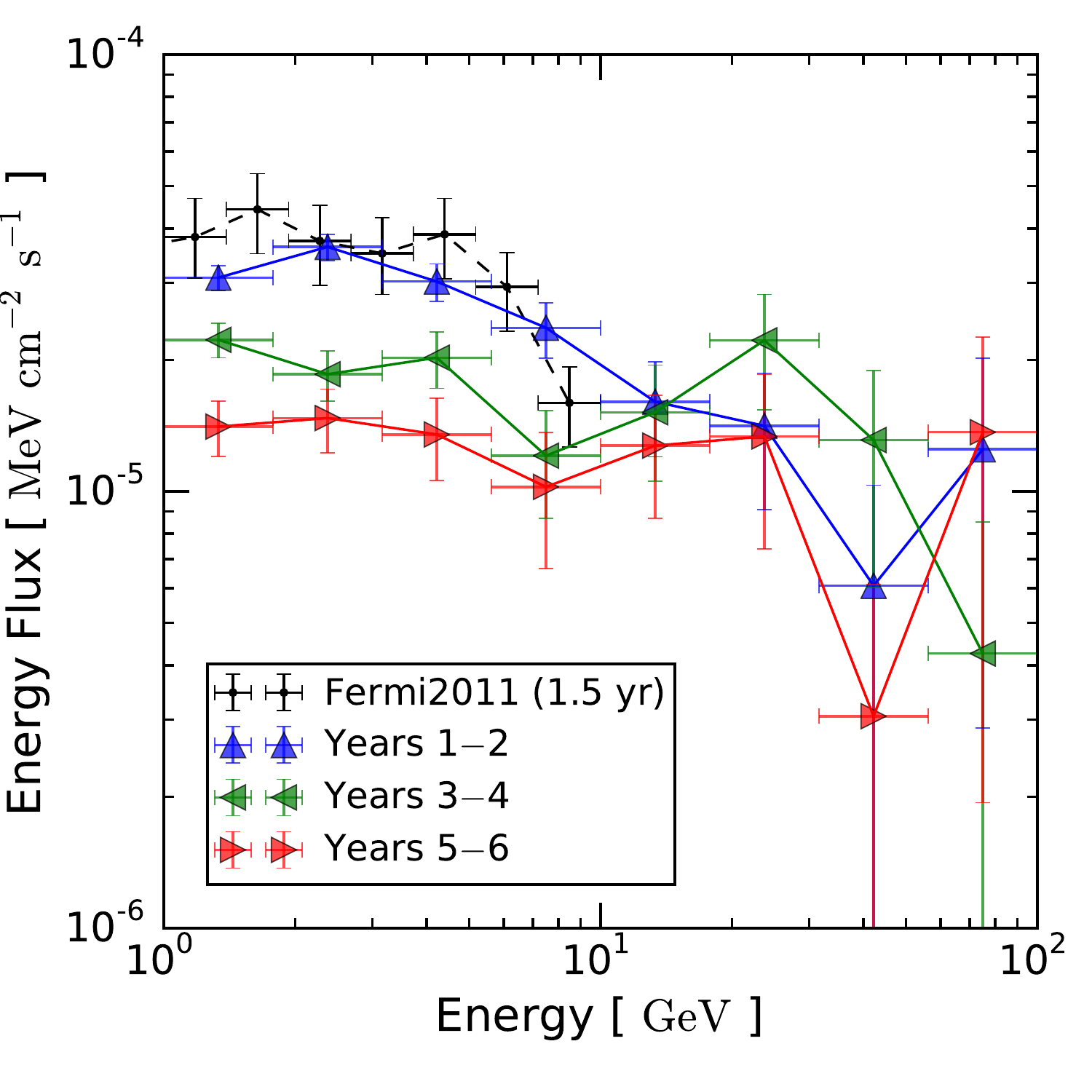}
\caption{Energy spectrum of the solar-disk flux, separated into three periods, each of two years.  The solar disk flux from first two years is consistent with Fermi2011, while the 1--10\,GeV data shows a significant reduction in later periods.}
\label{fig:2year_result}
\end{figure}
%+++++++++++++++++++++++++++++FIGURE++++++++++++++++++++++++++++++++++++++++%

\subsection{Solar-Disk Flux Spectrum} \label{sec:analysis}

We use a multicomponent fit  to extract the solar-disk component.  This exploits the facts that the Sun is spatially concentrated~(see Sec.~\ref{sec:fermi_prospects} for discussion on resolving the Sun),  the IC component is extended with a characteristic profile, and the diffuse background is spatially uniform.  The angular information allows us to fit the components individually for each energy bin, without requiring any assumptions about the energy spectra. 

We divide the Sun ROI into angular bins that are concentric rings of 1.5$^{\circ}$ width.  This choice is guided by the PSF of Fermi-LAT, which is $0.8^{\circ}$ at $1$\,GeV~(68\%).  Because the PSF improves above 1\,GeV and flattens out by $\sim10$\,GeV, the 1.5$^{\circ}$ bin ensures that the solar-disk component is always fully contained in the first angular bin. This criterion significantly simplifies the analysis.
Moreover, our choice of the uniform 1.5$^{\circ}$ angular bin across all energies is conservative.  The PSF of Fermi-LAT improves at high energies, so in principle one can afford a smaller angular bin at higher energy bins.  However, we expect the improvement from such an analysis will be small, given that the diffuse background is small.  For simplicity, we use constant angular bins across all energy range. 

With this angular binning, the distribution of the gamma-ray flux in the Sun ROI is modeled independently for each energy bin, as follows: 
\begin{eqnarray}
s_{i} &=& s_{1}\, \delta_{i 1}\,  \\
\nonumber b^{\rm IC}_{i} &=& f^{\rm IC} \, \sum_{j}{\cal E}_{i,j}\, \alpha^{-1}_{i,j}\,   \\
\nonumber b^{\rm BKG}_{i} &=& f^{\rm BKG} \, \sum_{j}{\cal E}_{i,j} 
\end{eqnarray}
where $s_{i}$, $b^{\rm IC}_{i}$, and $b^{\rm BKG}_{i}$ are the modeled photon counts for the solar-disk signal, as well as the IC and diffuse backgrounds in angular bin $i$.   ${\cal E}_{i, j}$ is the exposure for a pixel $j$ in bin $i$~(with unit $\left[ \rm cm^{2}\,s\,sr \right]$), and $\alpha_{i,j}$ is the angular distance from the center to a pixel $j$ in bin $i$.  
The solar-disk component is described by a Kronecker delta function, $\delta_{i1}$, which indicates that the first angular bin {fully contains} the solar-disk flux.  
The IC component is described by a normalization factor, $f^{\rm IC}$, times the total exposure weighted by $\alpha^{-1}$.  At the region of the solar disk~($\alpha< 0.27^{\circ}$), the IC component is strongly suppressed due to the anisotropy of the solar radiation and the occultation of the Sun~\cite{Orlando:2006zs, Moskalenko:2006ta, Orlando:2013pza}; we set the IC component to be zero in this region accordingly. 
The diffuse background component is radially isotropic, so it is only a normalization factor, $f^{\rm BKG}$, weighted by the total exposure.  

For each energy bin, we perform a profile likelihood analysis~\cite{Rolke:2004mj, Cowan:2010js}.  The likelihood function is a function of the signal parameter, $s_{1}$, and the nuisance parameters, $f^{\rm IC}$ and  $f^{\rm BKG} $:
\begin{equation}
{\cal L}(s_{1}; f^{\rm IC}, \, f^{\rm BKG}) = G(f^{\rm BKG})\prod_{i} P(s_{i} + b^{\rm IC}_{i}+ b^{\rm BKG}_{i} | d_{i}) , 
\end{equation}
where $P$ is the Poisson probability for the model to yield the observed number of photons, $d_{i}$. The product is taken over all angular bins.  
The Gaussian term, $G(f^{\rm BKG})$, constrains the diffuse background from deviating too much from the value determined from the fake-Sun method.  We take the variance of the Gaussian to be 20\% of the combined fake-Sun flux estimate, and assume that it is uncorrelated between energy bins.  The 20\% systematic uncertainty conservatively combines the 10\% variations among the individual fake Suns and the 10\% difference we observe from our fake-Sun method compared to that from Fermi2011.  The best-fit diffuse background normalization in the Sun ROI is found to be within 10\% of our fake-Sun estimate for all energy bins, which shows that the fake-Sun estimate is accurate and the choice of 20\% variance for $G(f^{\rm BKG})$ is conservative.  The normalization of the IC component is conservatively set as a nuisance parameter.  The final uncertainty of the extracted solar-disk component therefore includes the maximum normalization uncertainty of the IC component.

Figure~\ref{fig:profile} shows the angular distribution of the intensity in coarse energy bands, given by the number of photons in each angular bin divided by the total exposure.  The data points represent the total observed intensity with statistical error bars only, and the colored histograms represent the fit for the three individual components.  This simple model describes all features of the data well, and it is evident that the solar-disk component has a high signal-to-noise ratio.  

For each energy bin, we obtain the best-fit model parameters by maximizing the likelihood function with respect to all model parameters.  The uncertainty of the extracted solar-disk signal is found using the profile likelihood function, which is the likelihood function maximized over only the nuisance parameters.  Assuming the signal parameter is Gaussian-distributed, the 1-$\sigma$ error bar of the signal is determined by where the log-profile likelihood function differs from the best-fit value by $1/2$.  This uncertainty determination procedure is exact when the sample size is large, but is found to be reasonable for fairly small sample sizes~\cite{Cowan:2010js}.  We check explicitly that the log-profile likelihood function behaves close to the expected parabolic shape, which verifies the Gaussian-distribution assumption.  
In addition to the uncertainties estimated above, the gamma-ray flux has an overall 10\% systematic uncertainty from the effective area of the Fermi-LAT. 

We check our result using the same 1.5-year time period as in Fermi2011.  We find that our solar-disk component is consistent with that of Fermi2011, despite using different data sets~(\texttt{Pass 6} vs \texttt{Pass 7}), different energy and angular binning, different cuts, and a different analysis method.  This supports our analysis choices. 

For the full 6-year data set, we obtain a non-zero solar-disk signal in all eight energy bins from the likelihood analysis.  The detection significance can be estimated from the test statistic (TS  $\equiv2\Delta{\log\cal L}$), given by two times the difference between the best fit log-profile likelihood function and the one with the null hypothesis ~($s_{1} = 0$).  The Gaussian significance, to good approximation, is given by $\sqrt{\rm TS}$~\cite{Cowan:2010js}.  As a cross check, we obtain comparable best-fit parameters and uncertainties using a simple $\chi^{2}$ and $\Delta\chi^{2}$ analysis.

%---------------------------------------------------------Table--------------------------------------------------------------------------
\begin{table}
\caption{ For each energy bin, as defined, the total number of photons within 1.5$^{\circ}$ of the center of the Sun, the rounded best-fit number of photons due to the solar-disk signal, and the significance~($\sqrt{\rm TS}$) of the solar-disk flux detection. \label{tab:significance}    }
\begin{ruledtabular}
\begin{tabular}{lccc}
\hline
Energies~[GeV] 		& {Total cts.} 		& {Best-fit solar-disk cts.}		& $\sqrt{\rm TS}$	\\
\hline
1.0--1.8 			&	1468		&	961			&20.5			\\
1.8--3.2 			&	914			&	628			&17.7\\
3.2--5.6 			&	448			&	329			&13.6\\
5.6--10			&	188			&	133			&8.5\\
\hline
10--18 			&	92			&	67			&6.7\\
18--32 			&	55			&	42			&5.9\\
32--56 			&	16			&	10			&2.6\\
56--100 			&	12			&	7			&2.3\\

\end{tabular}
\end{ruledtabular}
\end{table}
%---------------------------------------------------------Table-----------------------------------------------------------------------

%+++++++++++++++++++++++++++++FIGURE++++++++++++++++++++++++++++++++++++++++%
\begin{figure*}[t]
%\centering
\includegraphics[angle=0.0, width=8.5cm]{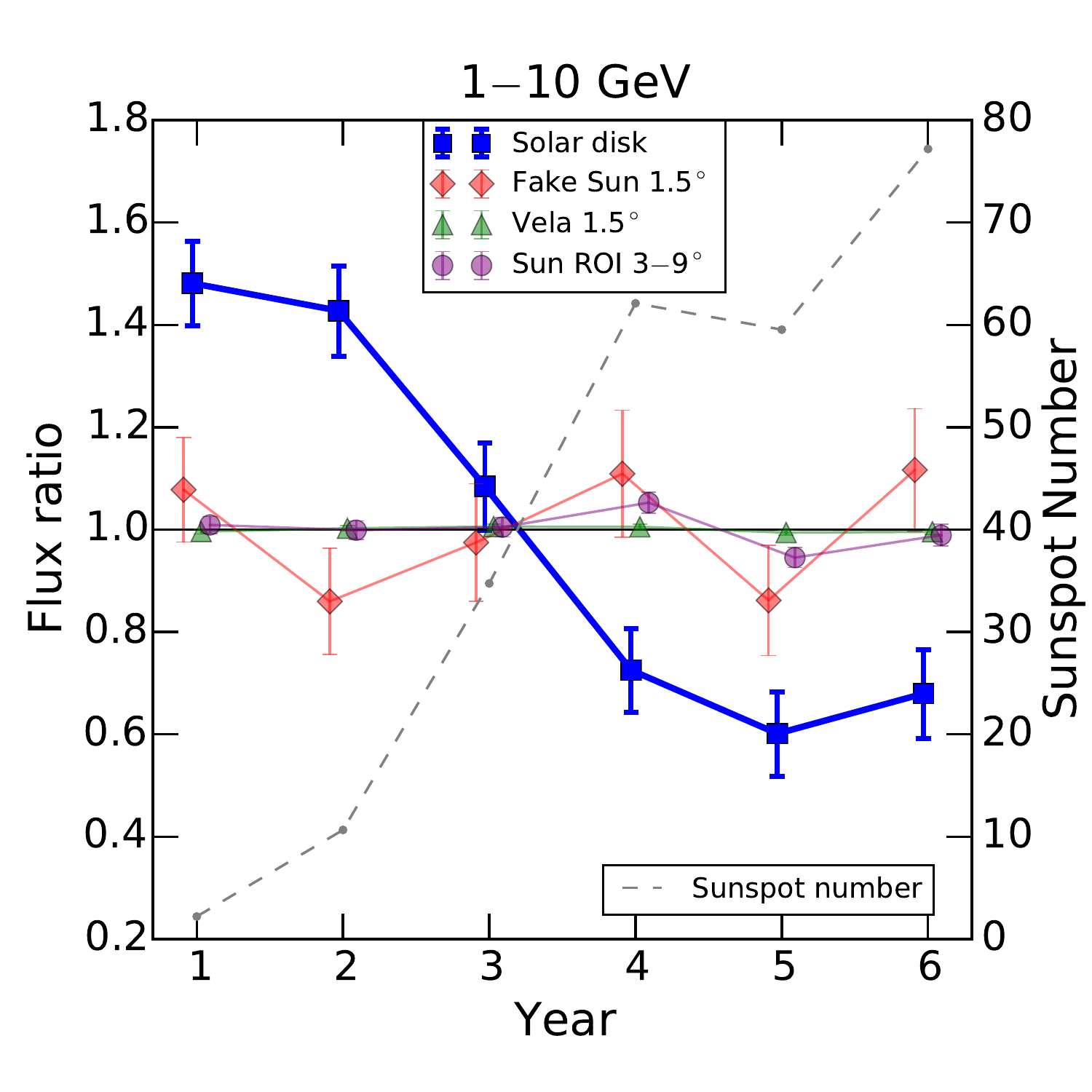}
\includegraphics[angle=0.0, width=8.5cm]{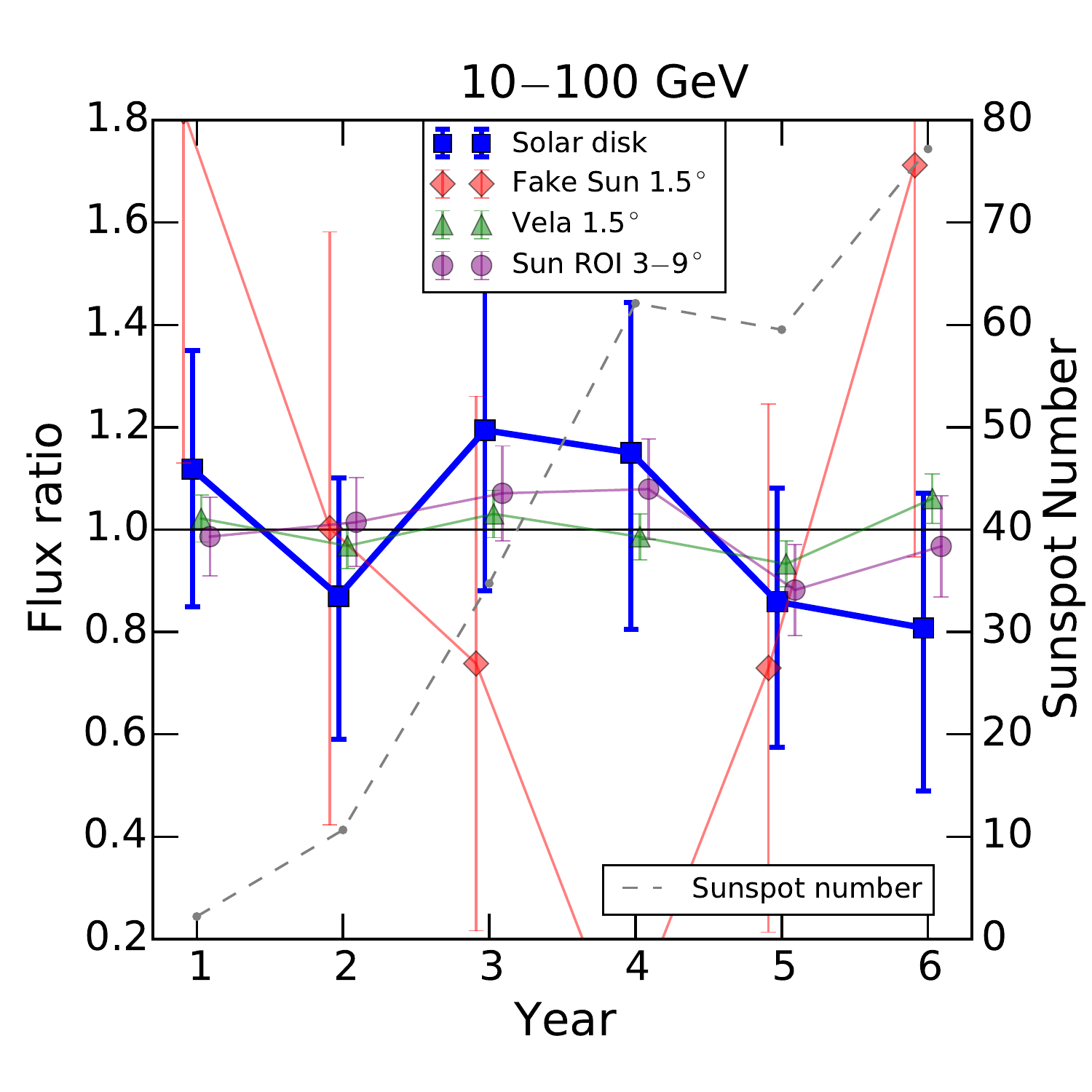}
\caption{{\bf Left:} For several sources, the ratio of the 1--10\,GeV flux in each year to its 6-year average (a time-independent source would fluctuate around unity).  The solar-disk component~(blue squares) demonstrates a clear decreasing trend and anticorrelation with the smoothed sunspot number, a tracer of solar activity.  Other sources (points displaced for clarity) should be and are consistent with being time-independent; see the text for details.  {\bf Right:}  Same, but for 10--100\,GeV.  No obvious trend is observed for the disk component, but the uncertainties are large.}
\label{fig:timemod}
\end{figure*}
%+++++++++++++++++++++++++++++FIGURE++++++++++++++++++++++++++++++++++++++++%

Table~\ref{tab:significance} summarizes our results, listing the energy bins, the total photon counts, and the best-fit numbers of photons in the solar-disk component, and  $\sqrt{\rm TS}$.  We find that the solar-disk component is significantly detected~($> 5 \sigma$) up to $\simeq 30$\,GeV, and is detected~($> 2 \sigma$) in each of the last two energy bins that go up to $100$\,GeV.  The lower detection significance at $>30$\,GeV is mainly due to not having enough statistics to distinguish the IC and solar-disk components.  We discuss the total solar gamma-ray flux more in Sec.~\ref{sec:wct}.

Figure~\ref{fig:data_result} shows the energy spectrum of the solar-disk component obtained in our 6-year analysis with 1-$\sigma$ error bars.  The spectrum extends to 100\,GeV without an obvious spectrum cutoff, though for energies $\gtrsim30$\,GeV, the error bars are large.  {The spectrum can be roughly described as a single power law,~$\propto E^{-2.3}$,} though the power-law fit is not particularly good.  For comparison, we also show the solar-disk component found in Fermi2011 and the {SSG1991 nominal model prediction on the solar-disk flux}, where in the former the error bars include systematics.  Comparing our result to that of Fermi2011, our analysis yields a similar spectrum with a lower normalization in the overlapping energy range.  We find that this is because the underlying flux has a significant time variation, as detailed in the next subsection.  

Compared to the central value of the SSG1991 prediction, our  1--10\,GeV result is still higher by a factor of about 5.  The flux normalization of the solar-disk gamma-ray flux remains am unsolved puzzle.  {To provide more context on the physical implications of this disagreement, we discuss and provide more details about the SSG1991 model in Sec.~\ref{subsubsec:ssg1991}.  }

\subsection{Time Variation of the Solar-Disk Flux}

Figure~\ref{fig:2year_result} shows the solar-disk gamma-ray flux energy spectrum obtained from our analysis when we divide the whole data set into two-year segments~(52 weeks per ``year'').  {In 1--10\,GeV, a decreasing trend in flux is clearly observed}.   Above 10\,GeV, the situation is unclear, due to the large error bars.  The time modulation of flux above GeV is already hinted at in Fig.~\ref{fig:time_curve}, where the 1--1.8\,GeV data showed a slow decline over the course of 6 years. 

To better quantify the time variation observed in 1--10\,GeV, we first combine the data into two broad energy bands~(1--10 and 10--100\,GeV), and then find the flux ratio for each energy band, which is the integrated flux in each year relative to that averaged over 6 years.  A time-independent source would fluctuate around unity.  

Figure~\ref{fig:timemod} shows the flux ratios in these two energy bands. In the 1--10\,GeV band, the solar-disk component demonstrates a significant time variation, an overall decreasing trend, in which the extremes differ by about a factor of 2 to 3.  {We estimate the statistical significance of the time variation by testing the data against the null hypothesis~(the underlying distribution is time independent) using a simple $\chi$-square test.  The $\chi$-squares are 104 and 1.6 for 5 degrees of freedom for the 1--10\,GeV 10--100\,GeV band, respectively.  This shows that the time-variation in the 1--10\,GeV band is highly significant, while the 10--100\,GeV data is consistent with being time independent. }

We note that in Fig.~\ref{fig:2year_result}, it can be seen that our 1--2 year result is compatible with the Fermi2011 spectrum in the overlapping energy range.  Given that flux only slightly decreases from the first year to the second year, this shows that our analysis with 18-months of data is compatible with that of Fermi2011. 

To make sure that the observed time variation is physical, we check the flux ratios of several gamma-ray sources as control samples.  First, we consider one of the fake Suns~(+180 days).  We find the total gamma-ray flux within 1.5$^{\circ}$ of its center, as in our solar-disk analysis.  This allows us to investigate possible fluctuations of the diffuse background.  For both energy bands, we find that they are consistent with being time independent.  Similar results are obtained when other fake Suns are used. 

Second, we consider the gamma-ray flux from the Vela pulsar~{(a constant gamma-ray source)}, which we use to validate our data selection procedure in Sec.~\ref{sec:data_selection}.  This allows us to check for unknown systematics in data selection.  The flux ratios of Vela demonstrate very small deviations from unity in both energy bands.  

Third, we consider the total flux in the 3$^{\circ}$--9$^{\circ}$ region from the Sun ROI, which allows us to check for peculiarities in the Sun ROI.  The flux ratios are again consistent with being time independent for both energy bands.  

None of the control samples demonstrates any systematic effects.  This means that the observed signal time variation is robust, and is a feature of the underlying gamma-ray production processes.  {This variation and its amplitude was never quantitatively predicted and this is the first time it is clearly observed.}

\subsection{Anticorrelation of the Solar-Disk Flux with Solar Activity}

We check whether the observed time variation is related to solar activity. 
Our analysis period coincides with solar cycle 24, which started with the solar minimum in 2009 and reached the solar maximum in  2014.  In Fig.~\ref{fig:timemod} we overlay the yearly smoothed sunspot number~\footnote{http://www.ips.gov.au/Solar/1/6}, which is a tracer of solar activity.  Though the sunspot number and the solar-disk gamma-ray flux vary with different amplitudes, the trends are clearly opposite.  In other words, the solar-disk gamma-ray flux anticorrelates with solar activity at least during the first half of the solar cycle 24. 

This trend is also qualitatively consistent with the EGRET observation.  The flux measured by Ref.~\cite{Orlando:2008uk} used data collected during 1991--1995, which is approximately the second half of solar cycle 22, when solar activity was declining from the solar maximum.  The anticorrelation explains the smaller flux observed by Ref.~\cite{Orlando:2008uk} compared to Fermi2011, who used data mainly from the solar minimum.    

Before this work, there was no direct evidence showing that the solar-disk gamma-rays are of cosmic-ray origin.  Though only rare solar flares are found to accelerate particles beyond 1\,GeV, it may be possible that some yet-unknown solar processes continuously accelerate particles up to the multi-GeV energy range.  However, one expects these solar processes would be correlated with solar activity, the opposite of the cosmic-ray framework~(detailed in the next section).  The anticorrelation with solar activity found in the solar-disk gamma-ray flux therefore strongly indicates that the bulk of the gamma-ray flux is induced by cosmic rays.  (Exploration of theoretical possibilities for the Sun itself to generate gamma rays that mimic the observed time variation is beyond the scope of this work.)

It is interesting to put the amplitude of this time variation into perspective, assuming the cosmic-ray production mechanism. The progenitors of 1--10\,GeV solar-disk gamma rays are $\sim$\,10--100\,GeV cosmic-ray protons.   The time variation~(or modulation) of the cosmic-ray flux at Earth is known to anticorrelate with solar activity, in the same sense as the solar-disk gamma-ray flux found in this work.  The cosmic-ray flux modulation at Earth is frequently described by the force field model with a single empirical parameter, the force field potential $\Phi$~\cite{1967ApJ...149L.115G, 1968ApJ...154.1011G, Cholis:2015gna}.  
The value of $\Phi$ can be extracted from precision ground-based neutron observations~\cite{2005JGRA..11012108U, 2011JGRA..116.2104U, phi:2015}. We obtain the corresponding values for our observation period by averaging over the monthly values.  In our analysis period, the maximum value of $\Phi$ was 630\,MV in 2014 and the minimum was 300\,MV in 2009. Taking these values, the maximum cosmic-ray flux is larger than the minimum by about 15\% at 10\,GeV and 2\% at 100\,GeV.  (For comparison, the extreme yearly values from 1964 to 2014 are about 1200\,MV and 270\,MV, which corresponds to about 50\% and 5\% difference in the cosmic-ray flux amplitude).  This is too small to explain the amplitude seen in Fig.~\ref{fig:timemod}.  This suggests that one needs additional modulation of the cosmic-ray flux in the inner solar system, variations in solar atmospheric magnetic fields that can affect cosmic rays of such high energies, or perhaps both to explain the observed variation amplitude. 

In fact, the Tibet air shower array found time variation in observations of $\sim$\,10\,TeV cosmic-ray shadows of the Sun.  During the solar maximum, the cosmic-ray shadows are shallower than during the solar minimum~\cite{2013PhRvL.111a1101A}.  This can be explained by coronal magnetic fields: cosmic rays are more severely deflected by the solar atmospheric magnetic fields during the solar maximum~\cite{2013PhRvL.111a1101A}.  This implies that it is more difficult for cosmic rays to go deep into the solar atmosphere during the solar maximum, which is consistent with our solar-disk gamma ray observations.  

The observation of time variation in the solar-disk gamma-ray flux therefore provides strong support for the cosmic-ray framework, which we discuss in detail in the next section.  

%%%%%%%%%%%%%%%%%%%%%%%%%%%%%%%%%%%%%%%%%%%%%%%%%%%%%%%%
%%%%%%%%%%%                                SECTION                                                           %%%%%%%%%
%%%%%%%%%%%%%%%%%%%%%%%%%%%%%%%%%%%%%%%%%%%%%%%%%%%%%%%%
\section{Theoretical Overview and Observational Outlook}
\label{sec:discussion}

In this section, we first review the cosmic-ray framework, i.e., how solar-disk gamma rays can be produced from cosmic-ray interactions with the solar atmosphere.  Experienced readers can skip the first part, and move on to the bulk of this section,  where we discuss some future prospects on solar gamma-ray theory and observations.  

\subsection{Physics of Solar-Disk Gamma Rays---The Cosmic-Ray Framework}
\label{subsec:physics}
\subsubsection{General Considerations}

The physics involved in the production of solar-disk gamma rays is complicated by the effects of magnetic fields.  To gain some physical insights, we describe some general cases and approximations, following SSG1991. 

Cosmic-ray propagation from the interstellar medium to the surface of the Sun is known to be affected by solar magnetic fields carried by the solar wind.  As a result, this propagation is also affected by solar activity~\cite{1967ApJ...149L.115G, 1968ApJ...154.1011G}.   Generally, cosmic rays with energy $\lesssim 10$\,GeV observed at the Earth are more suppressed when the Sun is more active.  Additional modulation of cosmic rays may occur when they propagate from the Earth to the Sun.  

Once cosmic rays reach the Sun, their motion is dominated by the magnetic fields in the corona and photosphere.  The Larmor radius of cosmic rays near the surface of the Sun sets a reference energy scale, $E_{c}$.  For cosmic-ray protons, taking the typical solar magnetic field strength, $B\sim$\,1\,G, and setting the Larmor radius, $L$,  to be the solar radius, ${ R}_{\odot} \simeq 7\times 10^{5}\,{\rm km}$, yields 
\begin{equation}
E_{c} \simeq 2\times10^{4}\,{\rm GeV} \left( \frac{L}{R_{\odot}} \right) \left( \frac{B}{1\,{\rm G}} \right). 
\end{equation}
A similar scale is obtained for sunspots, where the length scale is about $10^{3}$ times smaller, but the field strength is roughly $10^{3}$ times stronger.  The range of $E_{c}$ was estimated in SSG1991 to be between $\simeq 3\times 10^{2}$\,GeV and $\simeq 2\times 10^{4}\,{\rm GeV}$.  This scale separates the physics into three regimes: $E_{p}\gg E_{c}$, $E_{p}\ll E_{c}$, and $E_{p}\sim E_{c}$, where $E_{p}$ is the primary cosmic-ray energy.

When $E_{p} \gg E_{c} $,  one can ignore the magnetic fields. Cosmic rays and their interaction products travel in straight trajectories following the initial cosmic-ray momentum.  In this case, only gamma rays from the Sun limb are observable.  The Sun limb is the thin layer of the outer solar atmosphere that has high enough column density for cosmic rays to interact, but not so much that gamma rays are unable to escape.  This corresponds to a column density of $\cal O$(1) hadronic interaction length, which is similar to the photon absorption length. The Sun limb component is non-zero but is argued in SSG1991 to be small; it should also inherit the primary cosmic rays' spectral index~($\sim 2.7$).  The Sun limb component is expected to be time-independent. 

When $E_{p} \ll E_{c}$, cosmic rays propagate along solar atmospheric magnetic field lines.  Inward-pointing~(towards the Sun) cosmic rays are funneled into magnetic flux concentrations~(or flux tubes) in the photosphere, where the field strength is stronger and the matter density is higher.  Assuming adiabatic invariance, the inward-moving cosmic rays would be reflected by the magnetic field strength gradient~(magnetic reflection).  It is then possible for the cosmic rays to interact with the solar atmosphere on their way out and to produce gamma rays that point toward Earth. This mechanism, suggested in SSG1991, allows the whole solar disk to be involved in gamma-ray production, and thus enhances the flux.  Because the effects of magnetic fields on cosmic-ray propagation are energy dependent, the spectral index of the resultant gamma-ray flux could deviate significantly from that of the primary cosmic-ray spectrum.   During solar maxima, the strength of solar atmospheric magnetic fields increases~\cite{Vieira:2009uh}, so the magnetic reflection of cosmic rays are expected to occur at higher altitudes, where the density is lower.  This decreases the gamma-ray production efficiency during solar maxima compared to that during solar minima, which is qualitatively consistent with the time variation observed in this work. 

When $E_{p} \sim E_{c}$, no simple approximation can describe the physics.  The corresponding gamma-ray energy at $\sim 0.1 E_{c}$ marks the transition from the low-energy regime to the high-energy regime.  In other words, the gamma-ray flux, spectral index, and time-dependence should be intermediate between those of the two regimes above.  It is interesting to note that the robust detection of the solar-disk component at 30\,GeV and the non-observation of a spectral break in this work already requires that $E_{c}\gtrsim 300$\,GeV, which is close to the lower bound estimated by SSG1991. Interestingly, the result from the Tibet air shower array shows that cosmic rays at $\sim10$\,TeV are still affected by solar atmospheric magnetic fields~\cite{2013PhRvL.111a1101A}. 

\subsubsection{The SSG1991 Model}
\label{subsubsec:ssg1991}

We now briefly describe the SSG1991 ``naive" and ``nominal" cases for the solar-disk gamma-ray flux~(see Ref.~\cite{1991ApJ...382..652S} for details).  The SSG naive calculation ignores all the propagation and magnetic-field effects, assumes 100\% efficiency for cosmic-ray absorption in the solar surface, and counts all the gamma rays produced.  The naive case, therefore, is an robust theoretical upper limit on how much solar-disk gamma rays can be produced by cosmic rays.  It is not a physical model, and hence, it is not surprising that our flux and that from Fermi2011 is lower than this bound.  

The appropriate comparison with data is using the SSG nominal model, shown in Fig.~\ref{fig:data_result}.   In this case the cosmic-ray propagation was treated as a diffusion problem from the Earth to the Sun.  Primarily concerning the $E_{p} \ll E_{c}$ case, all cosmic rays were assumed to land on magnetic flux tubes, and then reflected with some efficiency.   With a chosen set of diffusion parameters, the cosmic-ray absorption rate was determined, which is roughly 0.5\%.  Finally, the magnetically enhanced gamma-ray flux was obtained by integrating the gamma-ray yield with the absorption rate and the path length distribution.  The upper edge of the green band in Fig.~\ref{fig:data_result} corresponds to the extreme case where all the cascade products are charged and contribute to the gamma-ray production.  The lower edge corresponds to the conservative case where all the cascade products are neutral, hence only primaries that interact after being reflected can contribute to the gamma-ray flux. These two cases bracket the theoretical uncertainty concerning the cascade development inside the flux tubes, but not other model ingredients.  

\subsection{Prospects for Solar-Disk Gamma-Ray Theory } 

As already discussed in Fermi2011, the SSG nominal model is unable to explain the observed gamma ray data. 
Our result, even if taken at solar maximum, is still inconsistent with the SSG nominal model. Therefore, it is necessary to revisit the modeling of the comic-ray framework.  Most likely, new implementations of cosmic-ray physics and solar physics are needed.  We will provide new theoretical investigations in our forthcoming papers. 

There are several key observations that the new model needs to address.  First, it needs to reexamine the effectiveness of solar magnetic fields in enhancing the gamma-ray flux at $E_{p} \ll E_{c}$.  In particular, SSG1991 estimated $\sim$\,0.5\% of the total available cosmic-ray energy at the Sun is converted to gamma rays, but observations suggest $\sim 5\%$, modulo the time variation.  Second, the high-energy gamma rays found in this work demand a proper treatment of the $E_{p}\sim E_{c}$ and $E_{p} \gg E_{c}$ regimes.  Third, the time variation found in this work, as well as that from the Tibet air shower array, show that the model should track the variations of solar magnetic activity.  Lastly, the model needs to quantitatively explain the observed amplitude of the time variation. 

With an accurate model of gamma-ray production, solar gamma-ray observations can be used to constrain model ingredients and parameters, thus providing a new probe of solar atmospheric magnetic fields and of cosmic-ray propagation in the solar system.  This is particularly promising given that many current and future instruments will have excellent sensitivity for continuously monitoring solar gamma rays.  

With a sufficient understanding of the solar-disk gamma rays, it will be possible to use the Sun as a laboratory to test new physics.  For example, a popular dark matter candidate is the Weakly Interacting Massive Particle~(WIMP), which can accumulate and annihilate in the core of the Sun after being gravitationally captured~(\cite{Krauss:1985ks, Silk:1985ax, Peter:2009mk}; see Ref.~\cite{Danninger:2014xza} for a recent review).  Typical WIMPs captured in the Sun generate negligible electromagnetic signals~\cite{Sivertsson:2009nx}.  However, non-minimal physics, such as inelastic dark matter~\cite{TuckerSmith:2001hy,Nussinov:2009ft, Menon:2009qj} and metastable mediators in the dark sector~\cite{Pospelov:2007mp, Finkbeiner:2007kk, ArkaniHamed:2008qn, Pospelov:2008jd}, can significantly enhance the electromagnetic signatures~\cite{Batell:2009zp, Schuster:2009au, Schuster:2009fc, 2011PhRvD..84c2007A}.   Understanding the standard model predictions is necessary to uncover or interpret any potential signatures from dark matter~\cite{Moskalenko:1991hm,1991ApJ...382..652S, Moskalenko:1993ke, Ingelman:1996mj, Hettlage:1999zr, Fogli:2006jk}.  For example, both the spectral information and time variation can be useful model differentiators.  We will further discuss the implications of high-energy solar observations for new physics in our forthcoming papers. 

\subsection{Prospects for the Inverse Compton Component}
In our analysis, the IC component is treated as a background.  However, with new data releases, which improve both statistics and data quality, a more precise study of the IC halo component is also warranted.  A minor tension between the data and the prediction for the IC component was found in Fermi2011, where the data seemed to be higher at 10\,GeV than expected.  A more precise measurement is needed to clarify the situation.  

A new study of the IC component will allow one to use gamma rays to probe the cosmic-ray electron density in the solar system~\cite{Moskalenko:2006ta}.  This is because the IC intensity is the product of the electron density and the photon density along the line of sight, with the latter being a known quantity.  
The IC component is therefore sensitive to electron densities from fairly close to the Sun to beyond the Earth's orbit.  In addition, if there is time variation in the IC component, its broad angular distribution may allow one to test the variation amplitude as a function of the distance to the Sun. 
These observations can help with understanding cosmic-ray modulation in the solar system, which despite many years of effort, is still under active investigation~\cite{Strauss:2012zza, Potgieter:2013pdj, Cholis:2015gna}.  This approach is complementary to solar-disk gamma-ray observations, which are strongly affected by the conditions of the solar atmosphere. 

Similar to our analyses, it is also interesting to characterize the IC component beyond 10\,GeV as well as search for long-term time variations.  Because point sources are not removed, our analysis is not optimized for the IC component.  With this caveat, we check the best-fit IC amplitude from our analysis, and we find no obvious time variation~(only $\sim$ 20\% scatter around the mean).  A more careful analysis is needed to provide a definitive statement. Analyzing the IC component is difficult at high energies, where statistics are low, and equally challenging at low energies for Fermi-LAT, where the PSF is $\sim$ 10$^{\circ}$ at 100\,MeV.  

\subsection{Prospects for Fermi and Future Space Missions}
\label{sec:fermi_prospects}

In this work, we use a straightforward analysis to characterize and robustly detect important features of the solar-disk gamma rays.  Future analyses and observations, with more optimized analysis procedures and improved data sets, can yield more precise measurements {or even discover new features}.  Below, we discuss some possible analysis improvements with Fermi. 

At high energies, where statistics are low, one can use an unbinned analysis to fully utilize the information carried by each photon.  In particular, better angular resolution at high energies may allow one to resolve the solar disk and locate hot spots~(as for solar flares~\cite{Ackermann:2014rma, 2014ApJ...789...20A}).  On the other hand, the improved angular resolution also means that the solar disk can no longer be treated as a point source.  One needs to take into account the fact that the astrophysical diffuse background and the IC component are reduced toward the solar disk~\cite{Orlando:2013pza}.  This also means that the one should avoid using the stacking procedures performed in this work, which slightly smears the position of the Sun according to the length of each time segment.  Instead, one should select the events and calculate the exposure in a solar-centric coordinate system.  

For improving statistics, one can potentially develop more optimized cuts.  For example, it is likely that the Galactic plane cut employed in this work can be improved, given that the Galactic plane gamma-ray intensity drops rapidly with latitude and can in principle be modeled.  This may improve the statistics by about a factor of 2.  In addition, the new Fermi data release, \texttt{Pass 8}~\cite{Atwood:2013rka}, has a larger effective area and better angular resolution.  Improving the statistics is particularly important for high-energy observations.  

Next-generation space gamma-ray telescopes can further improve the solar-disk observations in both time and energy range.  The apparent anticorrelation between the solar-disk gamma-ray flux and solar activity suggests that the flux should start to increase as we start to leave the solar maximum.  This can be checked with near-future data from Fermi.  Next-generation instruments, such as DAMPE~\cite{dampe}, GAMMA-400~\cite{Galper:2014pua}, and HERD~\cite{Zhang:2014qga}, will allow the Sun to be monitored at the GeV range even beyond Fermi's lifetime.  Though in principle Fermi is sensitive to gamma rays down to 10\,MeV, extracting the solar-disk signal is difficult due to the broad PSF.  Future missions such as PANGU~\cite{2014SPIE.9144E..0FW} and ComPair~\cite{Moiseev:2015lva} can provide improved sensitivity in the MeV range.  Low-energy observations could provide additional information on the time variation and probe potential leptonic components or even new solar-disk gamma-ray emission mechanisms.

\subsection{Prospects for Ground-Based Telescopes}
\label{sec:wct}

To expand solar gamma-ray observations into the TeV range and beyond,  large ground-based experiments are required.  It is impossible for air-Cherenkov telescopes to observe the Sun due the bright optical emission from the Sun itself.  The Sun, therefore, is a unique target for water-Cherenkov telescopes such as HAWC and LHAASO.

To assess whether water-Cherenkov telescopes can detect the Sun, we consider the total solar gamma-ray flux, including both the solar-disk and IC components.  We estimate this flux by finding the total flux within 1.5$^{\circ}$ of the Sun and subtracting the diffuse background.  In this case, the Sun is detected at $>5\,\sigma$ in all eight energy bins.  Assuming a single power-law spectrum, the total solar gamma-ray flux can roughly be described by $3.5\times 10^{-8}(E/{\rm GeV})^{-2.3}\,{\rm GeV^{-1}\, cm^{-2}\, s^{-1}}$ in 1--100\,GeV.

%+++++++++++++++++++++++++++++FIGURE++++++++++++++++++++++++++++++++++++++++%
\begin{figure}[t]
%\centering
\includegraphics[angle=0.0, width=8.5cm]{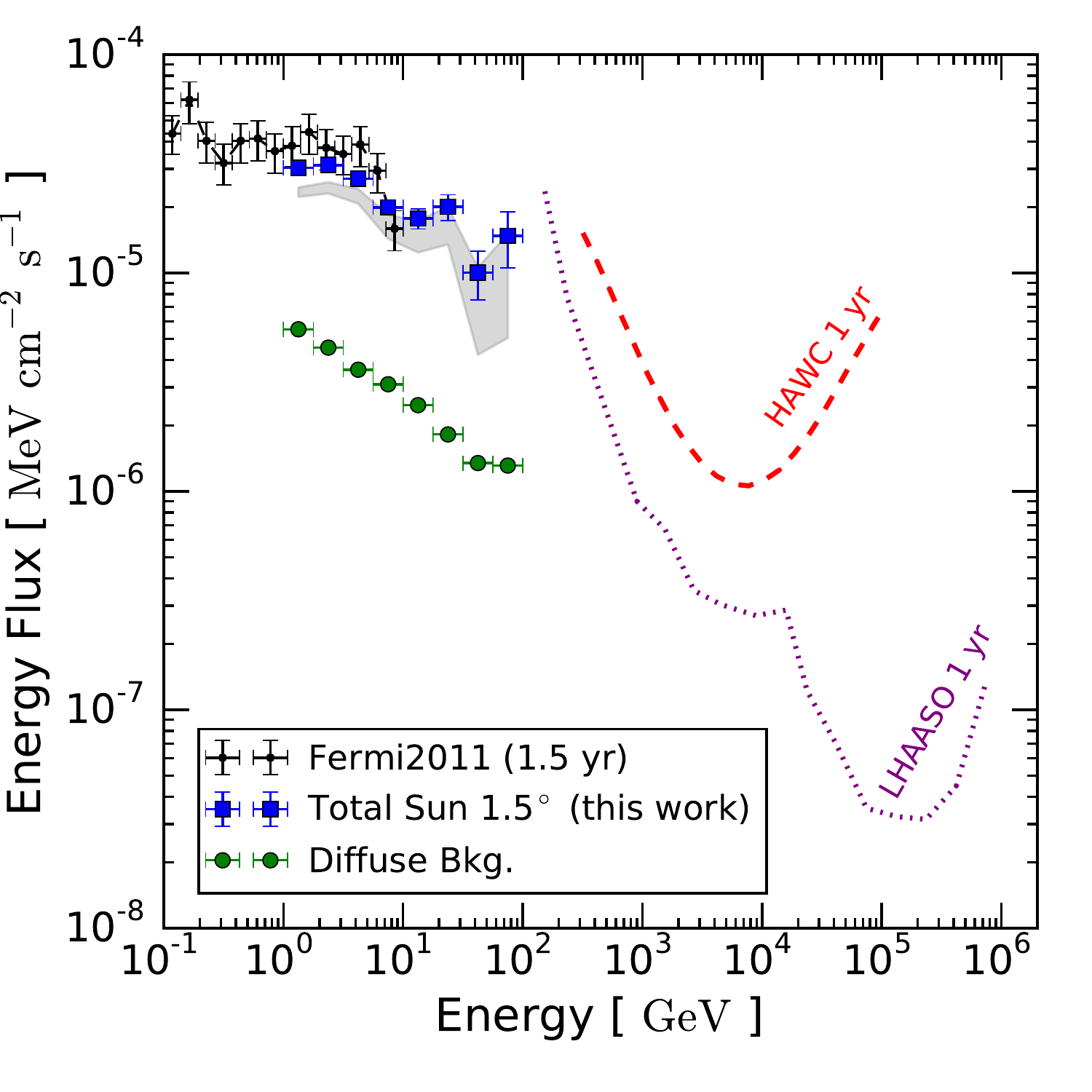}
\caption{ Energy spectrum of gamma rays from the Sun.  Blue squares are the total solar gamma-ray flux~(solar disk + IC) within $1.5^{\circ}$ of the Sun with only statistical error bars.  Black dots are the solar-disk-only component from Fermi2011.  The grey band shows the solar-disk-only component found in this work.  Green circles are the estimated diffuse background within $1.5^{\circ}$ of the Sun.  The differential point-source sensitivities for HAWC~\cite{Abeysekara:2013tza} and LHAASO~\cite{Zhen:2014zpa, hhh} are shown.    }
\label{fig:sens}
\end{figure}
%+++++++++++++++++++++++++++++FIGURE++++++++++++++++++++++++++++++++++++++++%

Figure~\ref{fig:sens} shows the total solar gamma-ray flux, the solar-disk-only component from Fermi2011, the solar-disk-only component found in this work, and the diffuse background within 1.5$^{\circ}$ of the Sun.  The total solar gamma-ray flux is clearly much larger than the diffuse background.  For comparison, we show also the sensitivity of HAWC~\cite{Abeysekara:2013tza} and LHAASO~\cite{Zhen:2014zpa, hhh}.  If the total solar gamma-ray flux follows the same spectral index to the TeV range, both HAWC and LHAASO should be able to detect the Sun.  

The water-Cherenkov telescopes are in a unique position to probe solar gamma rays.  In particular, they are sensitive to the $E_{p}\sim E_{c}$ and $E_{p} \gg E_{c}$ regimes.  Either a detection or an upper limit from the water-Cherenkov telescopes can provide valuable information on gamma-ray production from the Sun.

%%%%%%%%%%%%%%%%%%%%%%%%%%%%%%%%%%%%%%%%%%%%%%%%%%%%%%%%
%%%%%%%%%%%                                SECTION                                                           %%%%%%%%%
%%%%%%%%%%%%%%%%%%%%%%%%%%%%%%%%%%%%%%%%%%%%%%%%%%%%%%%%
\section{Conclusions}
\label{sec:conclusion}

Despite being the nearest star to us, much about the Sun's gamma-ray emission is still poorly understood. 
Previous study by the Fermi collaboration, who used 1.5 years of data, precisely detected the solar-disk gamma rays in 0.1--10\,GeV.  However, the flux is about ten times brighter than predicted.  Motivated by this puzzle, we focus on the solar-disk component, and use 6 years of public Fermi data to gain a better understanding of these gamma rays.  We employ a straightforward and conservative analysis to search for new features in the gamma-ray flux.  

Utilizing the improved photon statistics, we extend the observations to 100\,GeV.  As in Fermi2011, we find that the gamma-ray flux is higher than the central value of the SSG1991 prediction by about one order of magnitude in 1--10\,GeV, modulo time variation.  In addition, we detect the solar-disk component in 10--30\,GeV at $>$\,5\,$\sigma$, and in 30--100\,GeV at $>$\,2\,$\sigma$.  This is the first time the Sun is detected above 10\,GeV in gamma rays.  There are no theoretical predictions for solar-disk gamma rays in this energy range.  As a result, our observations demand further theoretical investigation.

Importantly, we find a significant time variation in the solar-disk gamma-ray flux over the analysis period, which apparently anticorrelates with solar activity.  This is the first clear observation of such a time variation, though it was hinted at in earlier studies~\cite{Orlando:2008uk, Abdo:2011xn}.   
This variation was not theoretically predicted, and its large amplitude deserve further investigation.  
Nonetheless, the anticorrelation with solar activity indicates that the bulk of the solar-disk gamma rays can be explained by cosmic-ray interactions in the solar atmosphere and the gamma-ray production process is strongly affected by the solar magnetic fields. 

Future observations with Fermi and other instruments may provide even more information about gamma rays from the Sun.  For example, the anticorrelation of the solar-disk gamma-ray flux with solar activity can be further confirmed with near-future Fermi data.  In addition, our robust detection~($> 5\,\sigma$) of the total solar gamma-ray flux shows that the Sun is a new and promising source for large water-Cherenkov gamma-ray telescopes, such as HAWC and LHAASO.  Observations from water-Cherenkov telescopes can provide important insights on the gamma-ray production processes in the TeV range.

This work lays the observational foundation for our future theoretical work, where we will investigate in detail how cosmic rays interact with the Sun under the influence of solar magnetic fields. We will study the multi-messenger signatures from these high energy processes, their implications for solar physics, cosmic-ray physics, and new physics. Gamma-ray studies of the Sun are still in their infancy, but have already yielded interesting results.  Future observations and the accompanying theoretical investigations may uncover even greater surprises.  

%%%%%%%%%%%%%%%%%%%%%%%%%%%%%%%%%%%%%%%%%%%%%%%%%%%%%%%%
%%%%%%%%%%%                                SECTION                                                           %%%%%%%%%
%%%%%%%%%%%%%%%%%%%%%%%%%%%%%%%%%%%%%%%%%%%%%%%%%%%%%%%%
\section*{Acknowledgments}
We thank Andrea Albert, Segev BenZvi, Brenda Dingus, Daniel Fiorino, Huihai He, Shunsaku Horiuchi, Yoshiyuki Inoue, Julie McEnery, Kohta Murase, Elena Orlando, Eric Speckhard, Andrew Strong, and especially Igor Moskalenko for helpful discussions.  {We appreciate the helpful comments from the anonymous referees, which improved the paper.}
KCYN and AHGP were supported by NASA grant NNX13AP49G awarded to AHGP and CR.  KCYN and JFB were supported by NSF Grant PHY-1404311 to JFB.  CR was supported by the Basic Science Research Program through the National Research Foundation of Korea funded by the Ministry of Science, NRF-2013R1A1A1007068.  

\bibliographystyle{h-physrev}
\bibliography{references}

\end{document}